\documentclass[12pt,preprint]{aastex}
\usepackage{lineno}
\usepackage{graphicx} \usepackage{amssymb} 
\usepackage{amsmath}
\usepackage{amsfonts}

\shorttitle{ANTARES point source search with 4-year data}

\shortauthors{Adri\'an-Mart\'inez et al.}

\begin{document}

\newcommand{\degree}{$^{\circ}$}
\newcommand{\dt}{$\Delta t$}

\title{Search for Cosmic Neutrino Point Sources with Four Years of Data from the ANTARES Telescope}

\author{S.~Adri\'an-Mart\'inez\altaffilmark{1},
I. Al Samarai\altaffilmark{2},
A. Albert\altaffilmark{3},
M.~Andr\'e\altaffilmark{4},
M. Anghinolfi\altaffilmark{5},
G. Anton\altaffilmark{6},
S. Anvar\altaffilmark{7},
M. Ardid\altaffilmark{1},
T.~Astraatmadja\altaffilmark{8,a},
J-J. Aubert\altaffilmark{2}
% ...
}

\altaffiltext{1}{Institut d'Investigaci\'o per a la Gesti\'o Integrada de les Zones Costaneres (IGIC) - Universitat Polit\`ecnica de Val\`encia. C/  Paranimf 1 , 46730 Gandia, Spain.}
\altaffiltext{2}{CPPM, Aix-Marseille Universit\'e, CNRS/IN2P3, Marseille, France}
\altaffiltext{3}{GRPHE - Institut universitaire de technologie de Colmar, 34 rue du Grillenbreit BP 50568 - 68008 Colmar, France }
\altaffiltext{4}{Technical University of Catalonia, Laboratory of Applied Bioacoustics, Rambla Exposici\'o,08800 Vilanova i la Geltr\'u,Barcelona, Spain}
\altaffiltext{5}{INFN - Sezione di Genova, Via Dodecaneso 33, 16146 Genova, Italy}
\altaffiltext{6}{Friedrich-Alexander-Universit\"at Erlangen-N\"urnberg, Erlangen Centre for Astroparticle Physics, Erwin-Rommel-Str. 1, 91058 Erlangen, Germany}
\altaffiltext{7}{Direction des Sciences de la Mati\`ere - Institut de recherche sur les lois fondamentales de l'Univers - Service d'Electronique des D\'etecteurs et d'Informatique, CEA Saclay, 91191 Gif-sur-Yvette Cedex, France}
\altaffiltext{8}{Nikhef, Science Park,  Amsterdam, The Netherlands}
% ...
\altaffiltext{a}{Also at University of Leiden, the Netherlands}
% ...

\begin{abstract}
  
  In this paper, a time integrated search for point sources of cosmic
  neutrinos is presented using the data collected from 2007 to 2010 by
  the ANTARES neutrino telescope. No statistically
  significant signal has been found and upper limits on the
  neutrino flux have been obtained. Assuming an $E_{\nu}^{-2}$
  spectrum, these flux limits are at $1-10\times$10$^{-8}$ GeV
  cm$^{-2}$ s$^{-1}$ for declinations ranging from $-90^{\circ}$ to
  40$^{\circ}$. Limits for specific models of RX J1713.7-3946 and Vela
  X, which include information on the source morphology and spectrum,
  are also given.

\end{abstract}


\section{Introduction}

One of the main goals of the ANTARES telescope~\citep{detector} is
the detection of cosmic neutrinos and the identification of their
sources. Neutrinos only interact via the weak interaction and are
stable, making them unique probes to study the high-energy
universe. The production of high-energy neutrinos has been
proposed~\citep{halzen2, bednarek, stecker} for several kinds
of astrophysical sources in which the acceleration of hadrons may
occur. In the interaction of cosmic rays with
matter or radiation, charged pions are produced. In the decay chain of
pions, neutrinos are produced. The detection of neutrinos may give valuable
information on the origin of cosmic rays. It would also settle
the question of the hadronic versus leptonic mechanism in several
sources from which high-energy gamma rays have been
observed~\citep{berezhko}.

The best neutrino flux upper limits up to PeV energies for the Southern
hemisphere have been established by the ANTARES experiment using
2007-2008 data~\citep{pspaper}. In the present paper, this analysis
is extended by adding two more years
of data with the full configuration of twelve detection
lines. Furthermore, the information on the amount of light produced in the events, which is
a quantity correlated to the neutrino energy and which helps to distinguish the
atmospheric neutrino background from a potential high-energy signal,
is taken into account.

The structure of this paper is as follows.  In
Section~\ref{sec:antares} the ANTARES detector is briefly described.
Sections~\ref{sec:data}~and~\ref{sec:simulation} describe the online
selection and the simulation, respectively. The track reconstruction
is explained in Section~\ref{sec:reco}. The selection of events is
described in
Section~\ref{sec:selection}. Section~\ref{sec:performance} is devoted
to the evaluation of the detector performance.  The search method and
the limit setting are described in Sections~\ref{sec:method}
and~\ref{sec:pes}. Section~\ref{sec:hits} shows how the search is
improved by including the energy information. Results
are presented in Section~\ref{sec:results}.  A cross-check based on an
alternative method is explained in Section~\ref{sec:em}. Finally, the
conclusions are summarised in Section~\ref{sec:conclusions}.

\section{ The ANTARES detector }
\label{sec:antares}

 The operation principle of neutrino telescopes is based on the detection of the
 Cherenkov light induced by relativistic muons produced in the
 charged current (CC) weak interactions of high-energy neutrinos close or
 inside the detector. The information on the time and position of the
 detected photons is used to reconstruct the muon trajectory, which is
 correlated with the direction of the incoming neutrino. Other
 signatures are also possible, such as the cascades produced in the CC
 interactions of electron and tau neutrinos and in the neutral current
 interactions of all neutrino flavours. In this analysis muons induced
 by high-energy neutrinos are used. For these events, the detector
 acceptance is large due to the long muon range and the neutrino direction is
 derived with an accuracy of a fraction of a degree.

\begin{figure}
 \begin{center}
\includegraphics[width=1.0\linewidth]{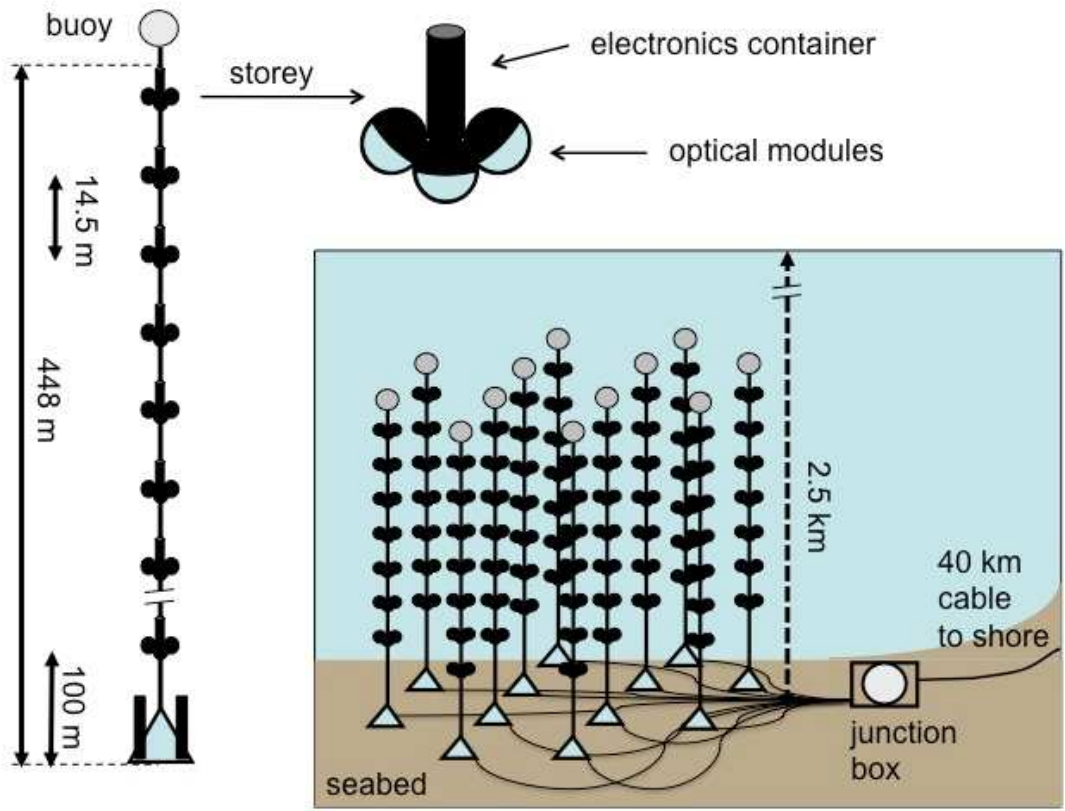}
\caption{ Schematic view of the ANTARES detector, consisting of
  twelve mooring lines connected to the shore station through an
  electro-optical cable.}
\label{fig:detector}
 \end{center}
\end{figure}

 The construction of the ANTARES detector~\citep{detector} was
 completed in 2008, after several years of site
 exploration and detector R\&D~\citep{oms, pmts, timing}.
 The detector is located at (42\degree 48' N, 6\degree 10' E) at a
 depth of 2475~m, in the Mediterranean Sea, at 40~km from the French
 town of Toulon. It comprises a three-dimensional array of 885 optical
 modules (OMs) looking 45\degree~downwards and distributed
 along 12 vertical detection lines. An OM~\citep{oms} consists of a ten-inch
 photomultiplier (PMT) housed in a glass sphere together with its base,
 a special gel for optical coupling and a $\mu$-metal cage for magnetic
 shielding. The OMs are grouped in 25 triplets (or storeys) on each
 line, except for one of the lines on which acoustic devices are
 installed~\citep{amadeus} and which therefore contains only 20
 optical storeys.  The total length of each line is 450~m, which are
 kept taut by a buoy located at the top of the line. The lower 100~m
 are not instrumented.  The distance between triplets is 14.5 m and the
 separation between the lines ranges from 60 to 75~m. The lines are
 connected to a central junction box, which in turn is connected to
 shore via an electro-optical cable. Figure~\ref{fig:detector} shows a
 schematic view of the detector.

 The detector also includes several calibration systems. The lines
 slowly move due to the sea current (up to $\sim$ 15~m at the top
 of the line in case of currents of 20~cm/s). A set of acoustic
 devices together with tiltmeters and compasses along the lines are
 used to reconstruct the shape of the lines and orientation of the
 storeys every two minutes~\citep{alignment}. The acoustic system
 provides the position of each optical module with a precision
 better than 15~cm. The time calibration is performed by means of a
 master clock on shore and a set of optical beacons (four along each
 line). This allows for a calibration of the time offsets of the
 photomultipliers with a precision better than 1 ns~\citep{timing}.

In this analysis, data from 29-1-2007 to 14-11-2010 are used. 
The total integrated live-time is
813 days, out of which 183 correspond to the five line period. Causes of
loss of efficiency are the periods of high bioluminescence and sea
operations. 

\section{ Online selection}  
\label{sec:data}

The charge and time information of all signals from the PMTs which
exceed a pre-defined threshold voltage, typically the equivalent to
0.3 single photo-electrons \citep{daq}, are first digitised into
'hits' and then sent to shore where they are filtered by a farm of
PCs.  For this analysis, two different filter algorithms were used to
select the events. Both are based on the assumption that the optical
background processes such as potassium-40 radioactive decays and
bioluminescence are not correlated and induce single photo-electron hits. 
Hence, a first selection of the signal requires
hits with a high charge (usually $>$ 3 photo-electrons) or coincident
hits within a time window of 20 ns on separate OMs of the same storey (L1
hits). The first trigger requires at least five L1 hits compatible
with a muon track in any direction. The second trigger is defined as
the occurrence of at least two L1 hits in three consecutive storeys
within a specific time window. This time window is 100 ns in case that
the two storeys are adjacent and 200 ns for next-to-adjacent
storeys. In addition to the events selected by the trigger, the
singles count rate of each OM is stored.

\section{ Simulations }
\label{sec:simulation}

Simulations are required for determining the acceptance and angular
resolution of the detector, since in the absence of a source these
quantities cannot readily be measured.

 The event simulation starts with the generation of upgoing 
 muon neutrino events using the GENHEN package \citep{baileythesis}, which uses 
 CTEQ6D \citep{cteq} parton density functions for computing the 
 deep inelastic charged current neutrino scattering cross 
 section. The events are weighted according to the cross section
 and their probability to survive the passage through the Earth.
 If the neutrino interaction occurs near the detector,
 the hadronic shower resulting from the break-up of the target
 nucleon is simulated using GEANT \citep{brunner}. Otherwise, only the 
 resulting muon is propagated to the detector using the MUSIC 
 code~\citep{music}.  
 Atmospheric muons reconstructed as upgoing are a source of background for a neutrino signal
 and their rejection is a crucial point in this analysis as will be described in Section \ref{sec:selection}. 
 Downgoing atmospheric muons were simulated with the program MUPAGE \citep{giada,yvonne} which provides 
 parametrised muons and muon bundles at the detector.
 
 Inside the detector, the Cherenkov photons emitted along a muon track and
 arriving on the OMs are simulated by sampling
 tabulated values of photon arrival times. 
 The arrival time distributions have been derived by tracking 
 individual photons taking into account the measured 
 absorption and scattering parameters \citep{watermodel}.
 
 The PMT transit time spread (TTS) is simulated by a Gaussian smearing of
 photon arrival times. Optical backgrounds are added to the events 
 according to the measured rates observed in the count rate data. 
 Similarly, simulated hits from inactive OMs are deleted from the event. 
 Sampling the count rate data from the runs selected for the analysis
 ensures that the simulation contains the same background and detector conditions
 as the analysed data set. The same trigger algorithms are applied to the 
 simulation and the data.

 An uncertainty of 50\% on the atmospheric muon flux is estimated using the same procedure described in \citet{agui2010}. For the
 atmospheric neutrino flux, a systematic uncertainty of 30\% is considered
 \citep{bartol_uncertainty}.
  
\section{ Reconstruction }
\label{sec:reco}

 Tracks are reconstructed from the hits in the triggered events
 using a multi-step algorithm (see \citet{thesis} for a more detailed
 description). The initial steps provide a starting point for the final
 maximisation of the track likelihood. The likelihood is defined as
 the probability density of the observed hit time residuals, $r$, given 
 the track parameters 
 (position at some arbitrary time and direction). The time 
 residual $r$ is defined as the difference between the
 observed and expected hit time for the assumed track parameters.

 It was found that the likelihood function has many local maxima 
 and that the maximisation procedure needs to be started with track 
 parameters close to the optimal values. 
 The initial steps in the algorithm provide this near-optimal 
 solution, estimating the track parameters using increasingly 
 refined score functions: a linear $\chi^2$ fit, a so-called
 'M-estimator' minimising $g(r)= \sqrt{ 1 + r^2 }$ and a 
 simplified version of the full likelihood fit. Each fit 
 uses increasingly more inclusive hit selections based 
 on the preceding stage.
 This sequence is started at nine different starting points 
 to further increase the probability of finding the global 
 optimum. 

 The final likelihood function uses parametrisations for the 
 probability density function (pdf) of the signal hit time residual, 
 derived from simulations. The pdfs also include hits arriving 
 late due to Cherenkov emission by  secondary particles or light 
 scattering. Furthermore, the probability  of a hit being due to 
 background is accounted for.  

 The quality of the track fit is quantified by the parameter 
\begin{equation}
 \Lambda \equiv \frac{\log(L)}{ N_{\rm hits}-5} + 0.1 \times (N_{\rm comp} -1 ),
\label{eq:lambda}
\end{equation}
which incorporates the maximum value of the likelihood, $L$, and the
number of degrees of freedom of the fit, i.e. the number of hits, $N_{\rm hits}$, used
in the fit minus the number of fit parameters; $N_{\rm comp}$ is the
number of times the repeated initial steps of the reconstruction
converged to the same result. In general, $N_{\rm comp}$ = 1 for badly
reconstructed events while it can be as large as 9 for well
reconstructed events.

The $\Lambda$ varible can be used to reject badly reconstructed
events, in particular atmospheric muons that are misreconstructed 
as upgoing. 
In addition, assuming that the likelihood function near the fitted
maximum follows a multivariate Gaussian distribution, the error on the
zenith and azimuth angles are estimated from the covariance matrix. 
The angular uncertainty on the muon track
direction, $\beta$, is obtained from these errors and can be used to further reject
misreconstructed atmospheric muons as discussed in Section~\ref{sec:selection}.

\section{Event selection}
\label{sec:selection}

 Neutrino candidates are selected requiring tracks reconstructed as
 upgoing and applying selection criteria. These criteria were chosen following a `blind' procedure, i.e. 
 before performing the analysis on data. The effect of the selection cuts on data, expected background and signal
efficiency are summarised in Table~\ref{tab:data}.
 
The estimated angular uncertainty on the muon track direction, $\beta$, is required to be smaller than 1 degree. 
This cut rejects 47\% of the atmospheric muons which are misreconstructed as upgoing tracks.

To further reject atmospheric
muons that were misreconstructed as upgoing, the quality variable $\Lambda$ is required to be larger than
-5.2. This value is chosen to optimise the discovery potential,
i.e. the neutrino flux needed to have a 50\% chance of discovering the signal 
at the 5$\sigma$ significance level assuming an $E^{-2}_{\nu}$ spectrum.  

\begin{figure}[!htp]
  \begin{center}
  \includegraphics[width=0.8\textwidth]{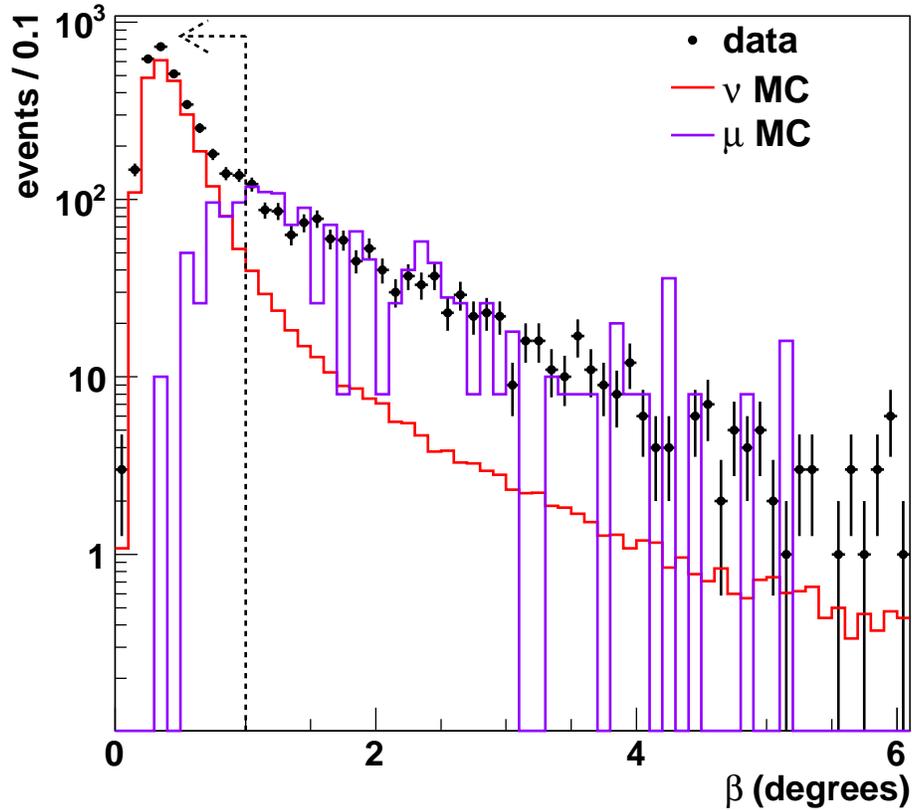}
  \caption{ Distribution of the estimate of
    the error on the direction of the reconstructed upgoing muon track after
    applying a cut on the quality variable $\Lambda > -5.2$. The red line shows the Monte Carlo atmospheric neutrinos, 
    the purple line the Monte Carlo misreconstructed atmospheric muons
    and the black dots the data. The vertical dashed
    line with the arrow shows where the selection cut is applied ($\beta < 1^{\circ}$). }
\label{fig:angerr}
\end{center}
\end{figure}

\begin{figure}[!htp]
  \begin{center}
    \includegraphics[width=0.8\textwidth]{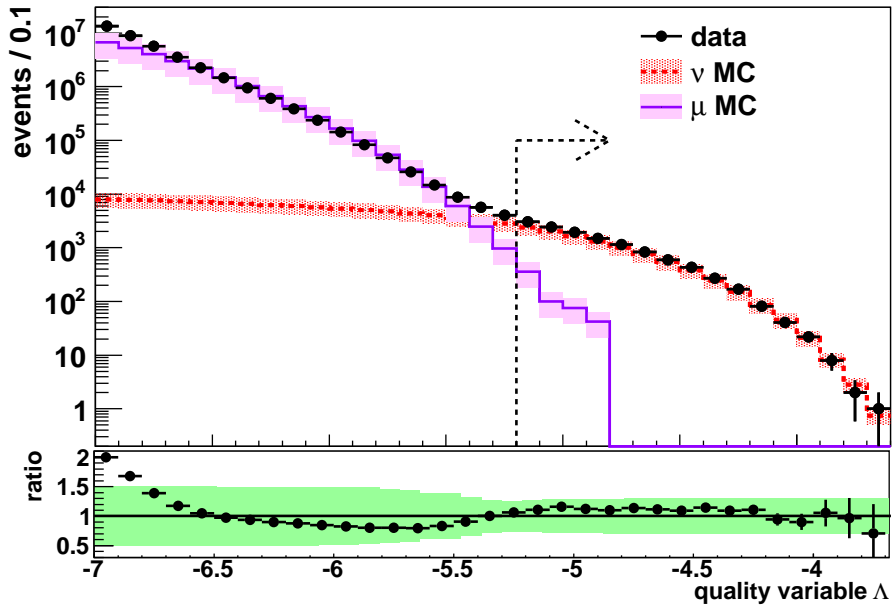}
  \caption{ Cumulative distribution of the reconstruction quality
    variable $\Lambda$ for upgoing tracks which have an angular error
    estimate $\beta <1^{\circ}$. The bottom panel shows the ratio between data
    and simulations. The red line is for Monte Carlo atmospheric neutrinos, 
    the purple line Monte Carlo misreconstructed atmospheric muons
    and the black dots the data. The vertical dashed
    line with the arrow shows where the selection cut is applied ($\Lambda > -5.2$). The purple and red
    bands show the systematic uncertainties on the simulations as explained in Section~\ref{sec:simulation}. The
    green band in the bottom panel shows the total contribution of these uncertainties.}
    \label{fig:lambda}
  \end{center}
\end{figure}

\begin{figure}[!htp]
\begin{center}
  \includegraphics[width=0.8\textwidth]{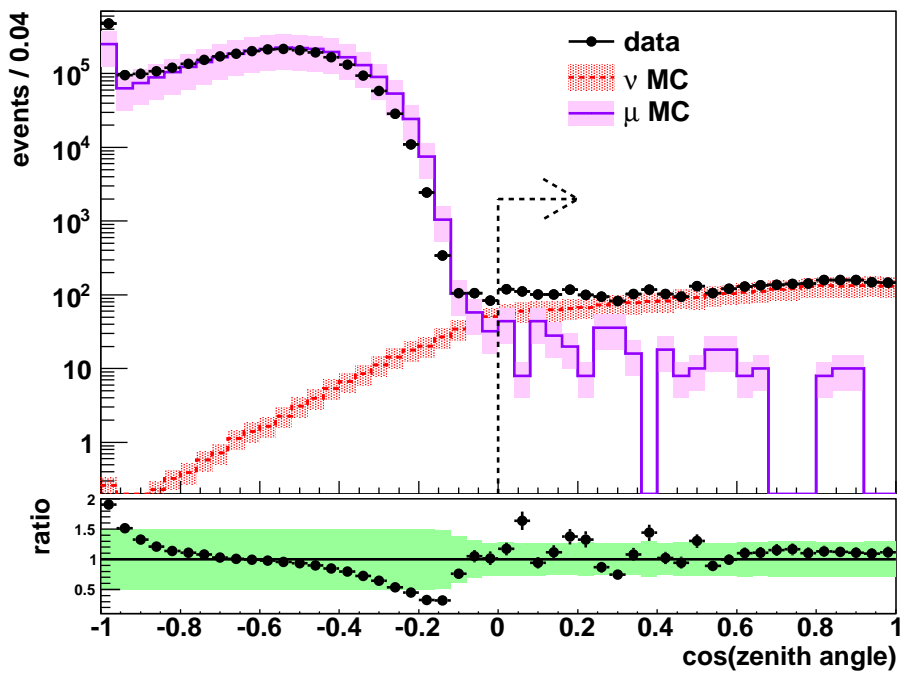}
\caption{ Distribution of the cosine of the zenith angle showing events with $\Lambda>-5.2$ and $\beta<1^{\circ}$. The bottom
  panel shows the ratio between data and the total contribution of simulations. 
  The simulated distributions are shown for atmospheric muons and neutrinos. Systematic uncertainties on Monte Carlo
  atmospheric muons and neutrinos are shown by the purple and red bands respectively. The green band corresponds to
  the sum of these uncertainties. The vertical dashed line with the arrow
  shows where the cut on the zenith angle is applied in order to select only upgoing events.}
\label{fig:zenith}
\end{center}
\end{figure}

Figure~\ref{fig:angerr} shows the distribution of $\beta$ for upgoing
events with $\Lambda > -5.2$.  The cumulative distribution of
$\Lambda$ for upgoing events is shown in Figure~\ref{fig:lambda}. The
cut on the angular error estimate $\beta$ is also applied. The excess
of data compared to simulations at the lowest values of $\Lambda$ is due to a non-simulated
contribution of events consisting of solely optical
background. Figure~\ref{fig:zenith} shows the distribution of the
reconstructed cosine of the zenith angle for both data and simulation.

The final sample consists of 3058 neutrino candidate events out of a
total of $\sim 4\times 10^{8}$ triggered events. Simulations predict
358 $\pm$ 179 atmospheric muons and 2408 $\pm$ 722 atmospheric
neutrinos, yielding a total expected events of 2766 $\pm$ 743. 
This is consistent with the observed rate within the quoted
uncertainties (see Section~\ref{sec:simulation}).

\begin{table}
 \centering
\begin{tabular}{c|l|llll}
\hline
\hline
 & data & atm. $\mu$ & atm. $\nu$ & $E^{-2}_{\nu}\nu$\\
\hline
Triggered events & $3.94\times 10^{8}$ & $3.06\times 10^{8}$ & $1.54\times 10^{4}$ & 100\%\\
\hline
Reco. upgoing events & $6.08 \times 10^{7}$ & $2.98\times 10^{7} $ & $1.24\times 10^{4}$ &  61\%\\
\hline
Reco. upgoing events + $\beta < 1^{\circ} $ & $3.90\times 10^{7}$ & $1.57\times 10^{7}$ & $8352$ & 44\%\\
\hline
Reco. upgoing + $\beta < 1^{\circ}$ + $\Lambda > -5.2$ & 3058 & 358 & 2408 & 23\%\\
\hline
\hline
\end{tabular}
\caption{ Number of events before and after applying selection cuts for data (second column) and 
	Monte Carlo simulations (third, fourth and fifth column). 
	The last column shows the percentage of signal events assuming a neutrino flux proportional to an $E^{-2}_{\nu}$ spectrum.}
\label{tab:data}
\end{table}

 \section{Detector performance}\label{sec:performance}
 The response of the detector to a neutrino signal proportional to an
 $E^{-2}_{\nu}$ spectrum was obtained using the simulation
 described in Section~\ref{sec:simulation} and applying the analysis cuts.

\subsection{Angular resolution}\label{sec:res}

\begin{figure}[!htp]
  \begin{center}
  \includegraphics[width=0.55\textwidth]{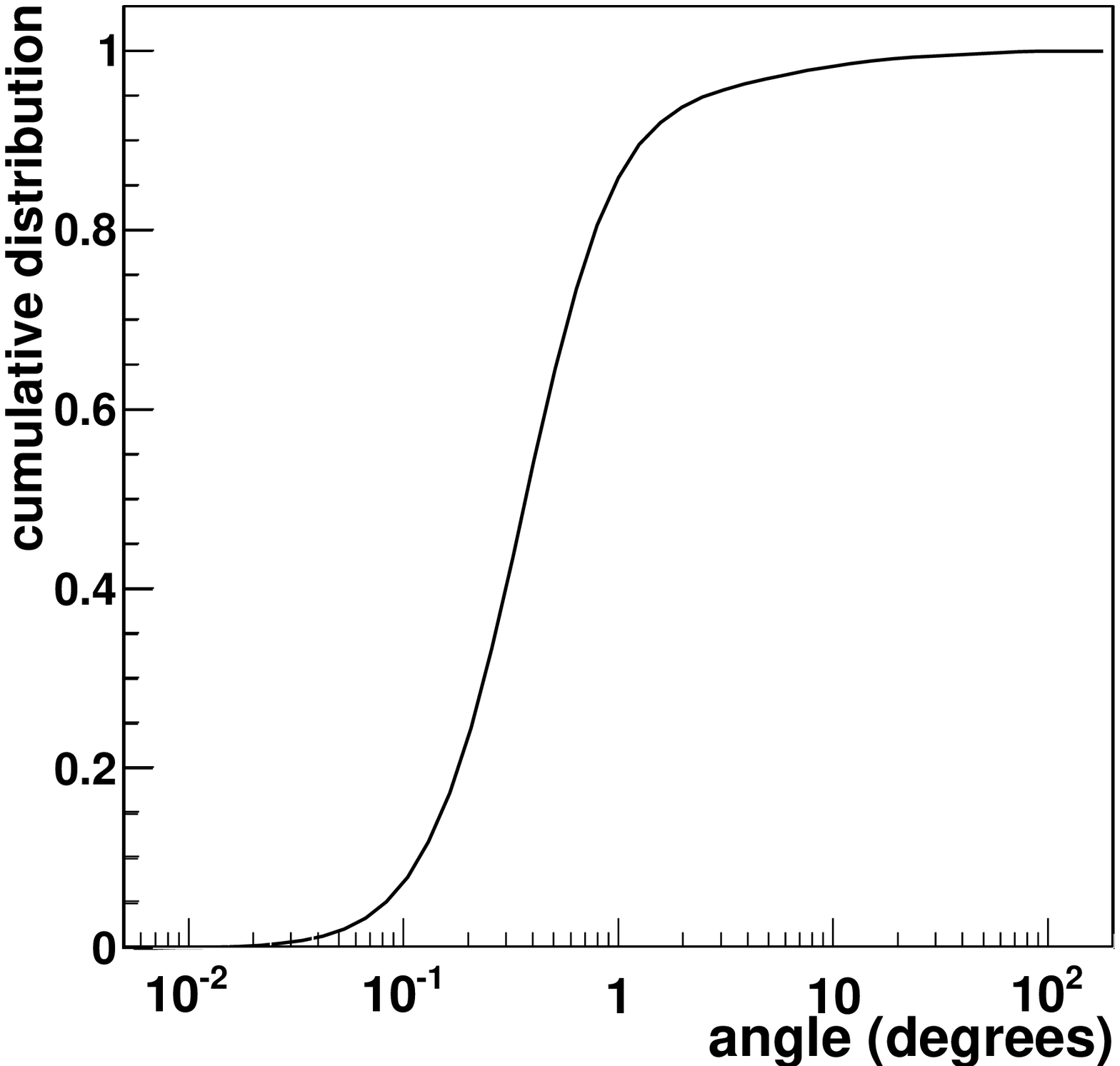} \\
  \includegraphics[width=0.60\textwidth]{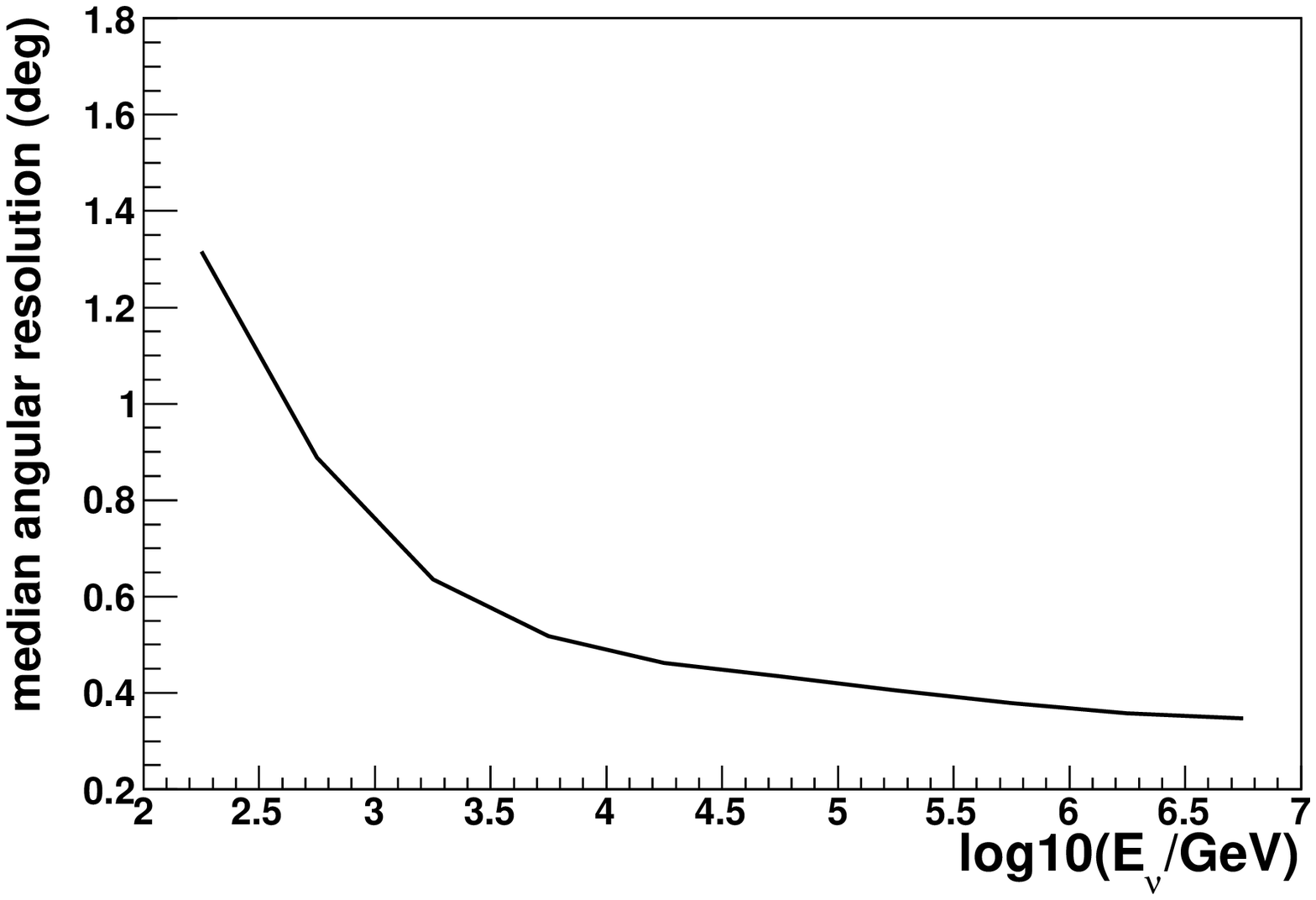}
  \caption{ (Top): Cumulative distribution of the angle between
    the reconstucted muon direction and the true neutrino
    direction for upgoing events of the whole data set. A neutrino spectrum proportional to $E_{\nu}^{-2}$ is
    assumed. (Bottom): The median angle as a function of
    the neutrino energy $E_{\nu}$.  
    In these plots the cuts $\Lambda > -5.2$ and $\beta < 1^{\circ}$ are
    applied.}
\label{fig:angres_periods}
\end{center}
\end{figure}

 Figure~\ref{fig:angres_periods} (left) shows the cumulative
 distribution of the angle between the direction of the reconstructed
 muon and that of the true neutrino. The median of this distribution is
 $0.46\pm 0.10$ degrees. Of the selected events, 83\% are
 reconstructed better than 1 degree. 
 For the data sample in which the
 detector was operational with all the 12 lines, the estimated angular
 resolution is $0.43\pm 0.10$ degrees.
 The median of this angular error for the full data set considered in the analysis is
 shown in Figure~\ref{fig:angres_periods} (right) as a function
 of the true neutrino energy.

 The systematic uncertainty on the angular resolution quoted above has
 been estimated by varying the hit time resolution, $\Delta t$, in the
 simulation.  Many possible effects can contribute to this resolution,
 including the PMT transit time spread, miscalibrations of the timing
 system and possible spatial misalignments of the detector. The hit
 time resolution directly impacts both the angular resolution and the
 number of events passing the quality criteria. Simulations using
 different $\Delta t$ values were compared with data in order to determine
 the best agreement in the lambda distribution. 
 and was obtained for \dt = 2.5 ns. This can be
 compared to the TTS of the PMT, which is 1.3~ns (standard
 deviation). However, the PMT time response is not Gaussian and the
 degraded resolution was found to partly account for the
 tails. A $\Delta t$ of 2.5 ns yields the quoted angular resolution of $0.46^{\circ}$
 and is the value used in the plots shown in this paper.

 For \dt =3.4 ns, the simulations show a deterioration in angular 
 resolution of 30\% and the number of selected neutrino events
 in data exceeds the simulated neutrino signal by 2$\sigma$, where 
 $\sigma$ refers to the uncertainty on the atmospheric neutrino flux model. 
 Hence, this value
 of $\Delta t$ is excluded by the data. Assuming a linear dependency, this argument translates to a 
 (1$\sigma$) uncertainty on the angular resolution of $\sim$ 15\%.
 
 The absolute orientation of the detector is known with an accuracy of $\sim 0.1^{\circ}$
 \citep{alignment}.

\subsection{Acceptance}\label{sec:acc}

\begin{figure}[!htp]
\begin{center}
  \includegraphics[width=0.60\textwidth]{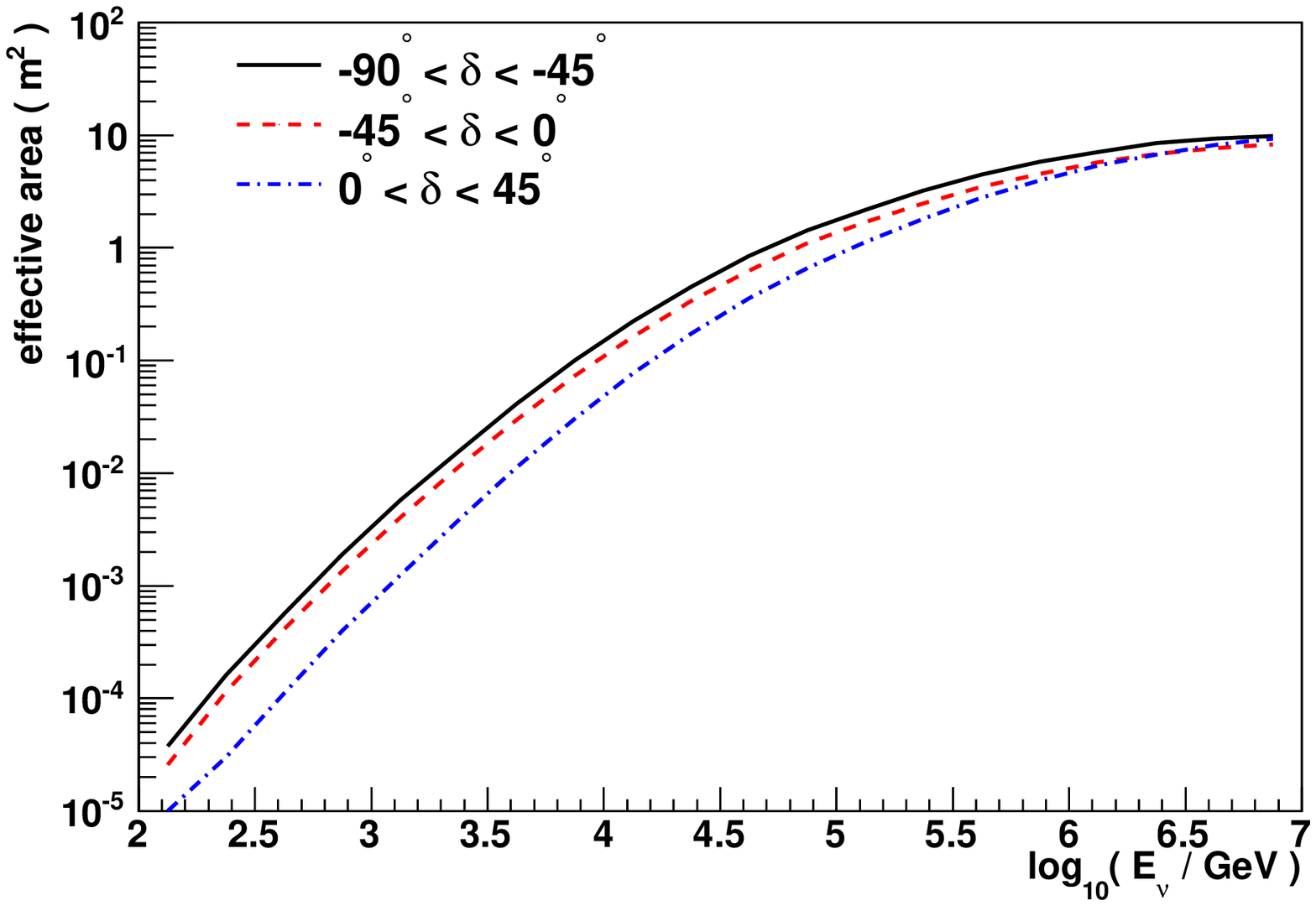}
  \includegraphics[width=0.60\textwidth]{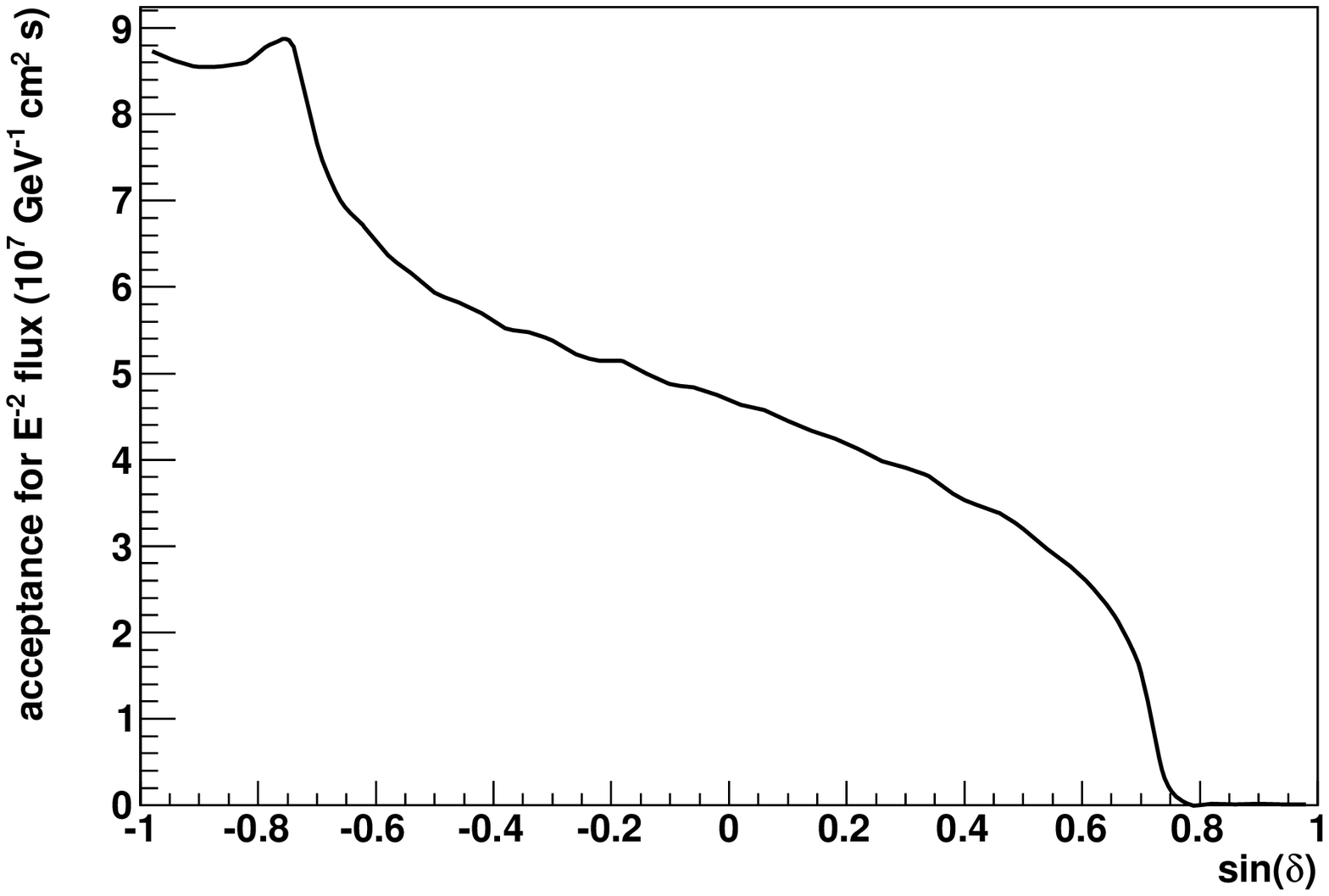}
\caption{ (Top): the neutrino effective area $A_{\nu}^{\rm eff}$ for the selected events as a function of the neutrino energy $E_{\nu}$ for three different 
declination bands. (Bottom): acceptance of the detector which is proportional to the number of events that would be detected and selected from a point-like source at a given declination assuming 
a flux of $10^{-7}\times (E_{\nu}/\rm GeV)^{-2}$ $\rm GeV^{-1}\rm cm^{-2}\rm s^{-1} $ as a function of the sine of the declination.}
\label{fig:acc}
\end{center}
\end{figure}

The neutrino effective area, $A_{\nu}^{\rm eff}$, is defined as the
ratio between the neutrino event rate, $R_{\nu}(E_{\nu})$, and the cosmic
neutrino flux, $\Phi(E_{\nu})$. The flux is assumed to consist of
equal amounts of $\nu_{\mu}$ and $\bar{\nu}_{\mu}$. The neutrino
effective area depends on the neutrino cross section, the absorption
of neutrinos through the Earth and the muon detection (and selection)
efficiency. It can be considered as the equivalent area of a 100\% efficient
detector. Figure~\ref{fig:acc} shows the effective area as a function
of the neutrino energy and declination, $\delta_{\nu}$.

The analysis is primarily concerned with cosmic sources emitting neutrinos with an $E^{-2}_{\nu}$ power law of the form
\begin{equation}
\frac{dN_{\nu}}{dE_{\nu}dtdS}=\phi \times \biggl(\frac{E_{\nu}}{\rm GeV}\biggr)^{-2} \rm GeV^{-1}\rm cm^{-2}\rm s^{-1},
\end{equation}
where the constant $\phi$ is the flux normalisation. The acceptance,
$A(\delta_{\nu})$, for such a flux, is defined as the constant of
proportionality between $\phi$ and the expected number of events in
the source direction and can be expressed in terms of the effective area as
\begin{equation}
A(\delta_{\nu}) = \phi^{-1}\int\int{dt dE_{\nu}A_{\nu}^{\rm eff}(E_{\nu},\delta_{\nu}) \frac{dN_{\nu}}{dE_{\nu}dt} }.
\end{equation}
The acceptance for this analysis is shown on Figure~\ref{fig:acc} (bottom). For a source at a declination of -90(0)$^{\circ}$, $A= 8.8 (4.8) \times 10^{7}$ $\rm GeV^{-1}$ $\rm cm^{2}$ $\rm s$ which means that a total of 8.8(4.8) neutrino candidates would be selected from a point-like source emitting a flux of $10^{-7} \times (E_{\nu}/\rm GeV)^{-2}$ $\rm GeV^{-1}$ $\rm cm^{-2}$ $\rm s^{-1}$.

To constrain the systematic uncertainty on the acceptance, a
comparison between the atmospheric neutrino data and a simulation was
performed, in which the efficiency of each of the OMs was reduced by
15\%, which yields a 12\% reduction of the signal events for an
$E_{\nu}^{-2}$ flux. The atmospheric neutrino yield would instead be
reduced by 40\% to be compared to the 30\% error on its flux
normalisation. Therefore, the $15\%$ uncertainty on the acceptance 
can be therefore considered a conservative choice.

\section{Search method}
\label{sec:method}
Two alternative searches for point-like neutrino sources have been
performed. The \textit{full-sky search} looks for an excess of signal events over the atmospheric muon and
neutrino background anywhere in the visible sky, i.e. in the
declination range [-90$^{\circ}$,+48$^{\circ}$]. In the
\textit{candidate list search}, the presence of an excess of events is
tested at the locations of the 51 pre-defined candidate sources given in
Table~\ref{table:clresults}. They include the 24 source
candidates from \citet{pspaper} and 27 new sources selected 
considering their gamma ray flux and 
their visibility at the ANTARES site as the selection criteria. Among the Galactic sources only TeV
gamma ray emitters are considered. No such requirement is imposed for
extragalactic sources as TeV gamma rays may be absorbed by the
Extragalactic Background Light \citep{ebl1, ebl2, ebl3}.

\begin{figure}[!htp]
\begin{center}
 \includegraphics[width=0.7\textwidth]{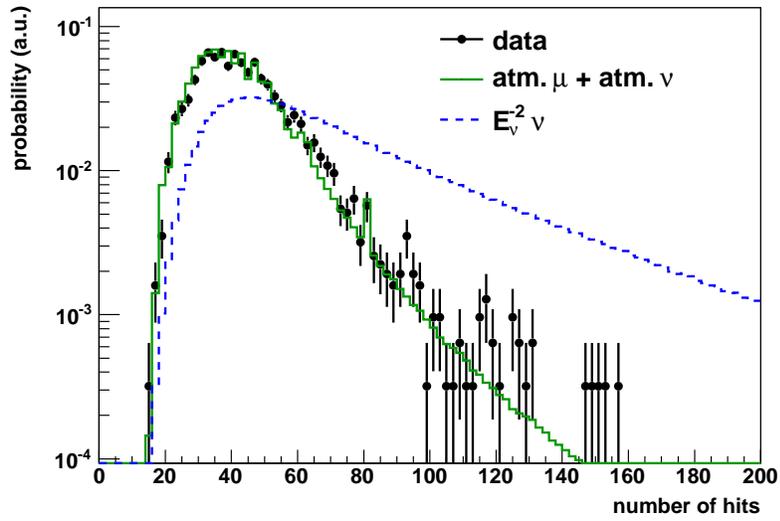}
 \caption{ Distribution of the number of hits used in the reconstruction, for 
        the selected data (black dots), and the total Monte Carlo background contribution, i.e. atmospheric muons 
        and atmospheric neutrinos (solid green line). The dashed blue line corresponds to the cosmic neutrino signal assuming an $E^{-2}_{\nu}$ spectrum.
        The distribution is normalised to the integral of the total number of events.
        All the 
        cuts described in Section~\ref{sec:selection} are applied.}
\label{fig:nhits}
\end{center}
\end{figure}

\begin{figure}[!htp]
\begin{center}
\includegraphics[width=0.7\textwidth]{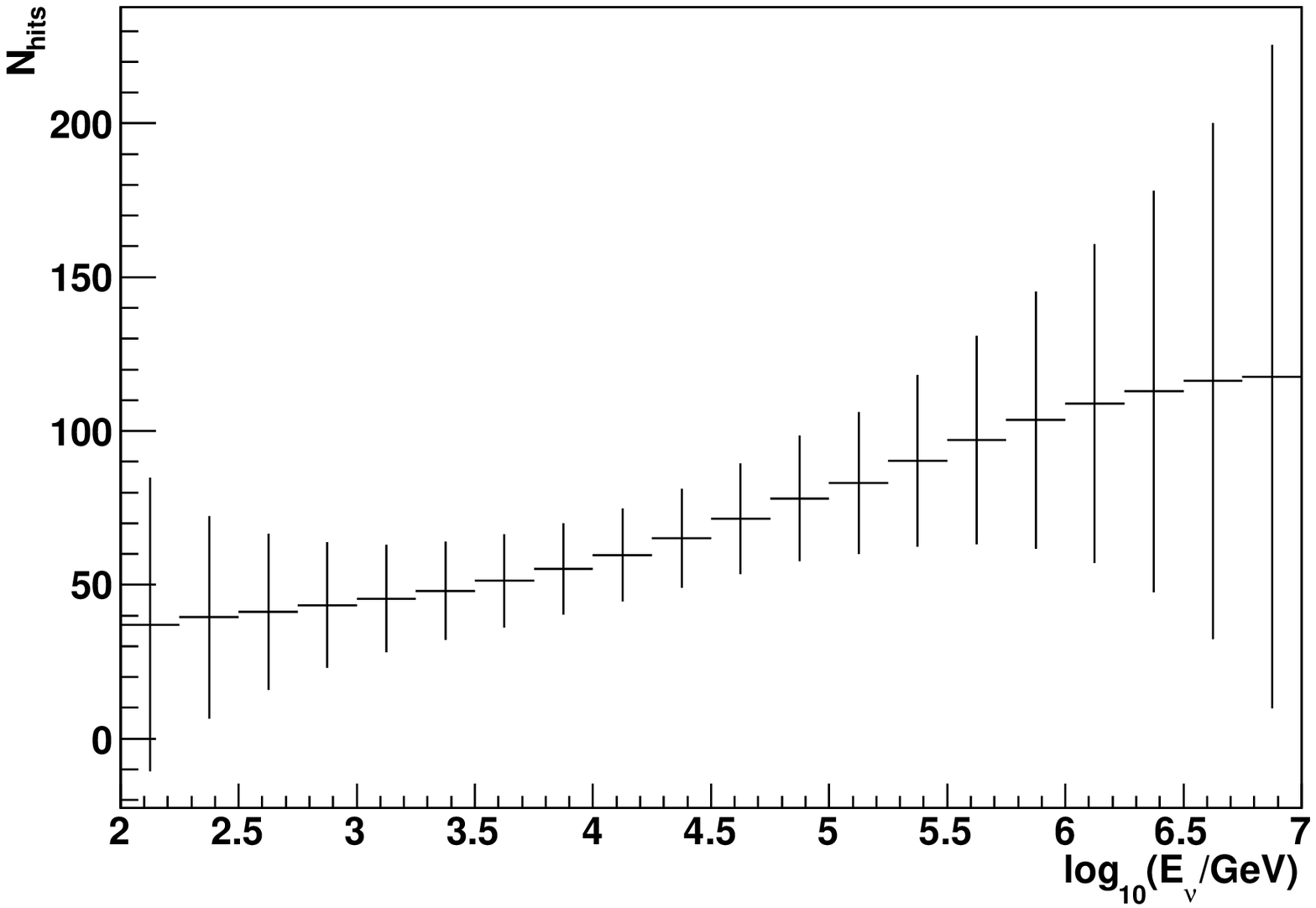}
\caption{ Distribution of the mean of number of hits, $\langle N_{\rm hits}\rangle$, as a function of $E_{\nu}$ for
 Monte Carlo signal events assuming a neutrino spectrum proportional
 to $E_{\nu}^{-2}$. The error bars represent the standard deviation of
 the $N_{\rm hits}$ distribution.
 The final cuts of the analysis described in Section~\ref{sec:selection} are included.}
\label{fig:nhits_profile}
\end{center}
\end{figure}

 The algorithm for the cluster search uses an unbinned maximum likelihood \citep{barlow} which 
 is defined as
\begin{equation}
\begin{split}
\log{\cal L}_{\rm s+b} &= \sum_{i} \log[\mu_{\rm s}\times {\cal F}(\psi_{i}(\alpha_{\rm s},\delta_{\rm s}))\times {\cal N}^{\rm s}(N_{\rm hits}^{i}) \\
&+{\cal B}(\delta_{i})\times {\cal N}^{\rm bg}(N_{\rm hits}^{i})]-\mu_{\rm s} - \mu_{\rm bg},
\label{eq:lik}
\end{split}
\end{equation}
where the sum is over the events; ${\cal F}$ is a parametrisation of
the point spread function, i.e. the probability density function of
reconstructing event $i$ at an angular distance $\psi_{i}$ from the true
source location $(\alpha_{\rm s},\delta_{\rm s})$; ${\cal B}$ is a parametrisation of the
background rate obtained from the distribution of the observed
declination of the 3058 selected events; $\mu_{\rm s}$ and $\mu_{\rm bg}$ are the mean
number of signal events and the total number of expected background events; 
$N_{\rm hits}^{i}$ is the number of hits
used in the reconstruction. ${\cal N}^{\rm s}(N_{\rm hits}^{i})$ and
$ {\cal N}^{\rm bg}(N_{\rm hits}^{i})$ are the probabilities of
measuring $N_{\rm hits}^{i}$ hits for signal and background
respectively. The distribution of $N_{\rm hits}^{i}$ for data 
and Monte Carlo events is shown in Figure~\ref{fig:nhits}. Figure~\ref{fig:nhits_profile}
shows the distribution of $N_{\rm hits}$ for signal as a function of the true neutrino energy.

In the candidate list search, the sum in Equation \ref{eq:lik} incorporates
the events located in a cluster within 20 degrees of the source. Events
further away have ${\cal F}\simeq 0$ and thus contribute a constant factor
to the likelihood.
In the full-sky search potentially significant clusters are first
identified by selecting at least 4 events within a cone of 3 degrees diameter.
Using a larger diameter or a bigger/lower number of required events increases the 
computation time without a significant improvement in the sensitivity.

In the candidate list search the likelihood is maximised by
numerically fitting the mean number of signal events, $\mu_{s}$, with the source location fixed. In the full-sky search the likelihood maximisation yields the source coordinates and $\mu_{s}$ for each cluster.
After likelihood maximisation a test statistic, ${\cal Q}$, is computed:
\begin{equation}
{\cal Q} = \log{\cal L}_{\rm s+b}^{\rm max} - \log{\cal L}_{\rm b}, 
\label{eq:test_q}
\end{equation}
where $\log{\cal L}_{\rm s+b}^{\rm max}$ is the maximum value of the likelihood provided by the fit and $\log{\cal L}_{\rm b}$ is the likelihood computed for the background only case 
($\mu_{s} = 0$). A large value of ${\cal Q}$ indicates a better compatibility with the signal hypothesis. In case of a full-sky search only the cluster with the largest value of 
the test statistic is considered.

\section{Pseudo-experiment generation and limit setting}
\label{sec:pes}
\begin{figure}[!htp]
\begin{center}
  \includegraphics[width=0.8\textwidth]{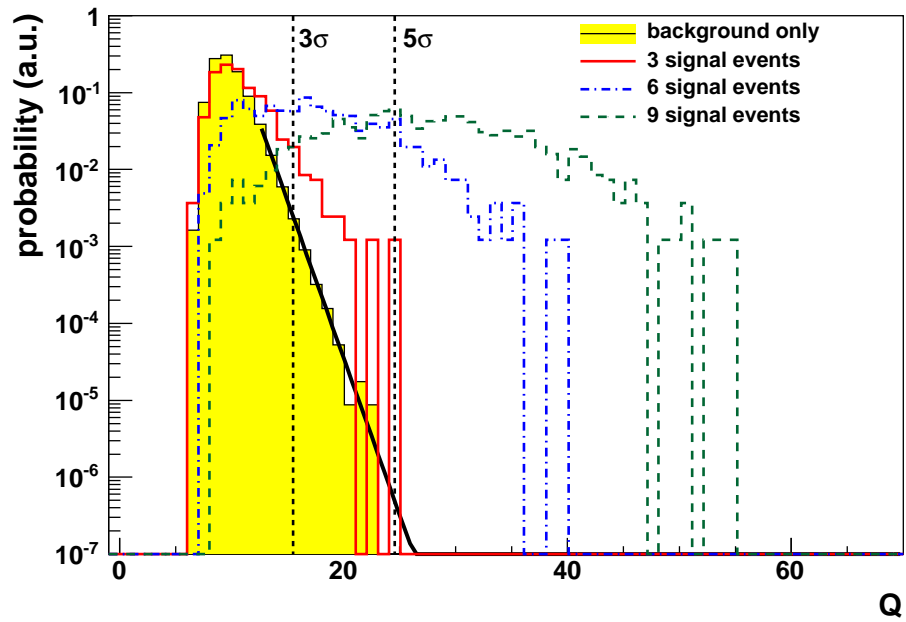}
\caption{ Distribution of the test statistic ${\cal Q}$ for the full-sky search. The full yellow histogram is for the background only experiments. The red, blue and green lines are for 3, 6 and 9 signal events generated from a source with an $E^{-2}_{\nu}$ spectrum at a declination of $-70^{\circ}$. The vertical dotted lines show the values of  ${\cal Q}$ corresponding to the 3$\sigma$ and 5$\sigma$ significance level.}
\label{fig:q}
\end{center}
\end{figure}

In order to evaluate the sensitivity of the analysis, pseudo-experiments are generated simulating background and signal. 
Background events are randomly generated by sampling the declination from the parametrisation 
${\cal B}$. The right ascension is sampled from a uniform distribution. The simulation of the signal is performed by adding events around the coordinates of the source, 
sampling in this case the angular distance $\psi_{i}$ of the event $i$ to the source location from a three-dimensional distribution of the reconstruction error as a function 
of the declination and the number of hits. At this stage the systematic uncertainties on the angular resolution and on the absolute orientation of the detector are included by varying the simulated parameters of the events of each experiment, such as the zenith and azimuth angles, with the required uncertainty. 
An example of the distribution of the test statistic ${\cal Q}$ obtained by performing the search on a large number of pseudo-experiments for the full-sky search is shown in Figure~\ref{fig:q} for the background only hypothesis and for experiments where several signal events are added to the background at a declination $\delta = -70^{\circ}$. 
Figure~\ref{fig:q} also shows the values of ${\cal Q}$ corresponding to p-values of 2.7 $\times$ $10^{-3}$ and 5.7 $\times$ $10^{-7}$, i.e. 3$\sigma$ and 5$\sigma$. To compute the latter value, the distribution of the test statistic for the background only hypothesis has been extrapolated using an exponential fit.

The median sensitivity and the flux upper limits at 90\% confidence level (CL) are computed following the Feldman \& Cousins prescription \citep{FC}. The systematic uncertainty on the acceptance is taken into account in the computation.

\section{Discovery potential}
\label{sec:hits}

\begin{figure}[!htp]
\begin{center}
  \includegraphics[width=0.7\textwidth]{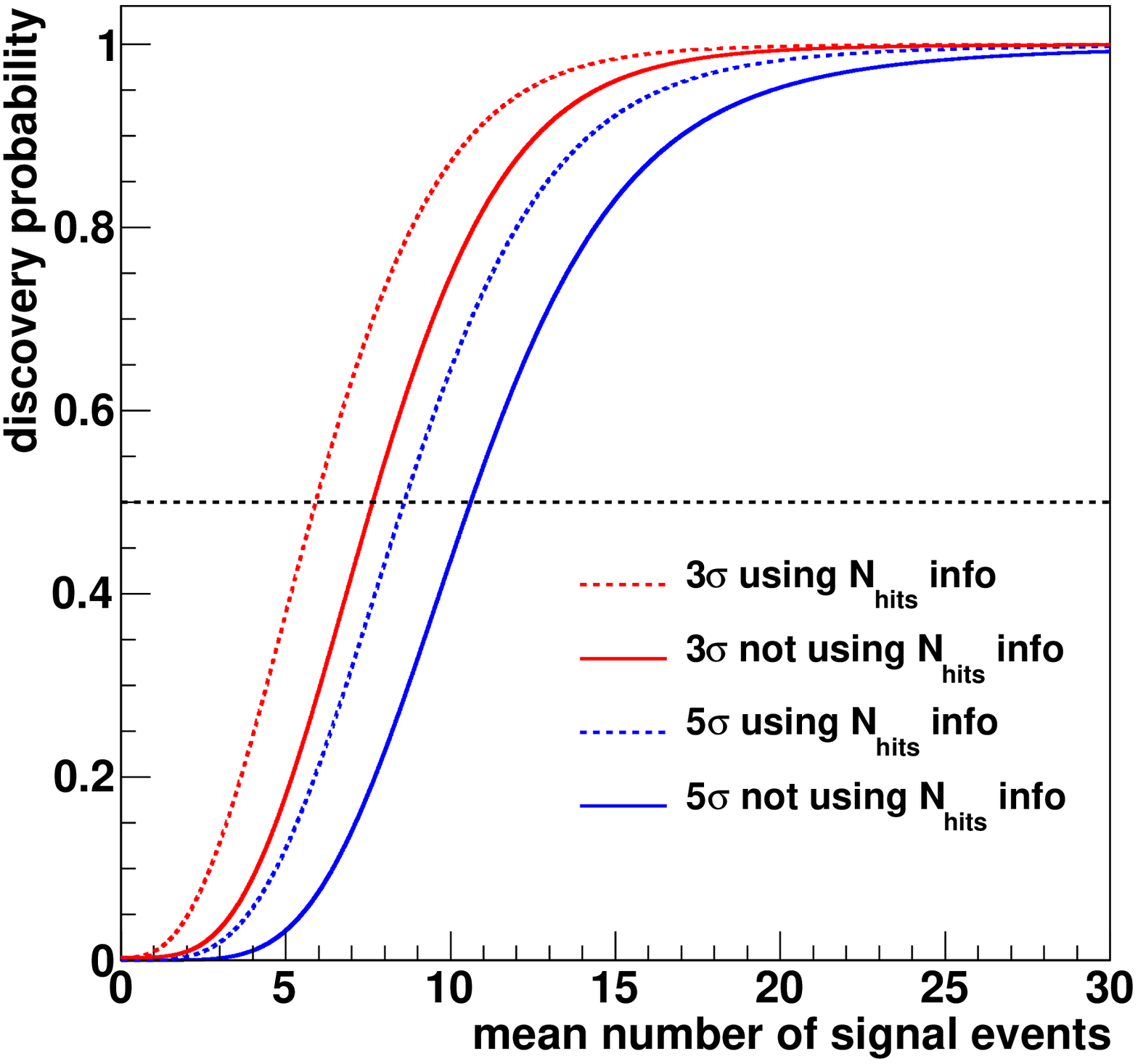}
  \caption{ Probability for a 3$\sigma$ (red lines) and 5$\sigma$ (blue lines)
    full sky search discovery as a function of the mean number of
    signal events from a source at $\delta=-70^{\circ}$ with a
    neutrino spectrum proportional to $E_{\nu}^{-2}$. The dotted blue
    and red lines are for the likelihood described, the solid lines refer to
    the case where $N_{\rm hits}$ is not used. The horizontal dotted black line corresponds to the
    probability to make a discovery in 50\% of the pseudo-experiments.}
\label{fig:disco}
\end{center}
\end{figure}

 Figure~\ref{fig:disco} shows the probability of making a discovery at the 
 3$\sigma$ and 5$\sigma$ significance level, as a function of the mean number
 of signal events. The same curves are shown for a search which 
 does not use $N_{\rm hits}$ in the likelihood. 
 The inclusion of the $N_{\rm hits}$ pdfs in the likelihood function reduces 
 the number of events (and therefore the signal flux) needed for a 
 discovery by $\sim 25\%$.
  
The worsening of the 3$\sigma$ and 5$\sigma$ discovery probability for a neutrino flux model with an exponential cut-off parametrised as $dN/dE = 
\phi\times(E_{\nu}/\rm GeV)^{-2}$ $exp(-E_{\nu}/E_{\rm c})$, with $E_{\rm c}$ the cut-off energy, was estimated.
In this case, for a source at a declination of $\delta=-70^{\circ}$, the mean number of signal events needed for a $5\sigma$ discovery assuming a cut-off energy $E_{\rm c}=1$ TeV is a factor 2 higher compared to that without an exponential cut-off.

Simulations show that for a source with Gaussian extension $\sigma_{\rm source} = 1^{\circ}$ at a declination of $\delta=-70^{\circ}$, the 
flux needed to claim a $5\sigma$ discovery is a factor 1.2 higher compared to a point-like source.  

\section{Results}
\label{sec:results}
\begin{figure}[!htp]
\begin{center}
\includegraphics[width=1.1\textwidth]{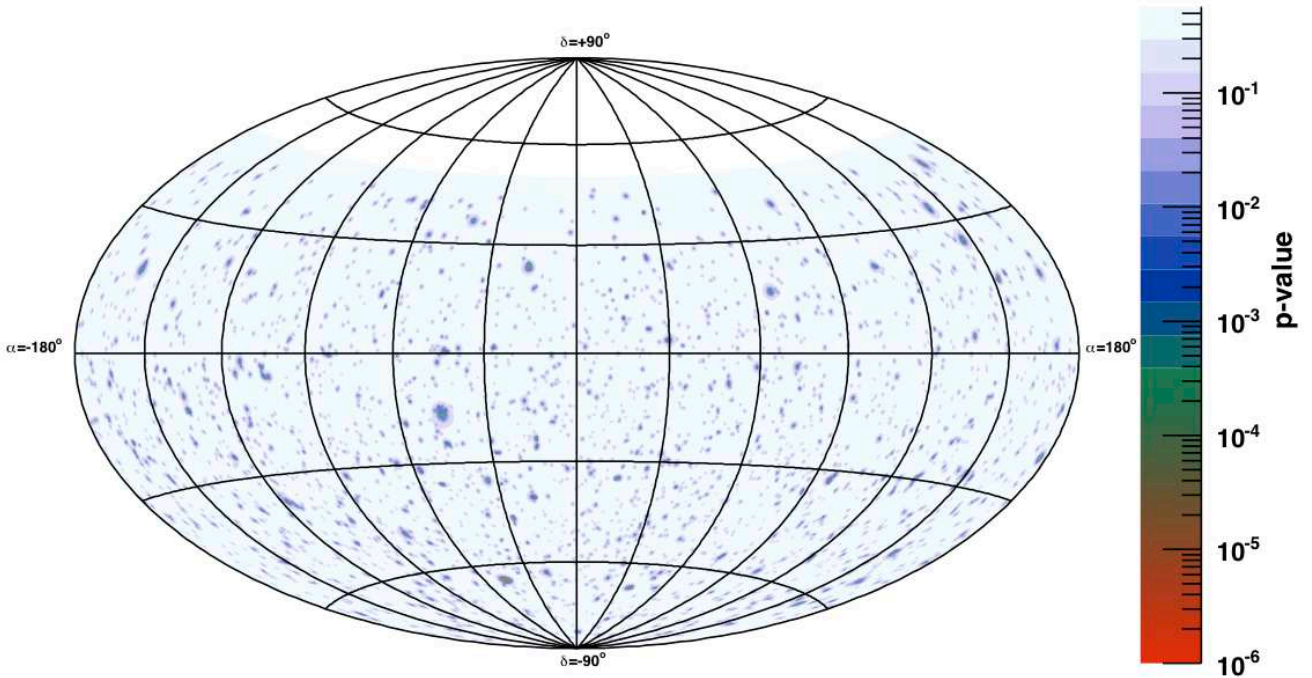}
\caption{ Sky map in equatorial coordinates showing the p-values obtained for the point-like clusters evaluated in the full-sky scan; the penalty 
factor accounting for the number of trials is not considered in this calculation.}
\label{fig:skymap}
\end{center}
\end{figure}

A map in equatorial coordinates of the pre-trial significances of every point in the sky that 
is visible below the horizon at the ANTARES site is shown in Figure~\ref{fig:skymap}. 

In the full-sky search the most significant cluster is located at
$(\alpha,\delta) = (-46.5^{\circ},-65.0^{\circ})$, where 5(9) events
are within 1(3) degrees of this position.
For this cluster the fit assigns 5.1 as signal
events, and the value of the test statistic is ${\cal Q}=13.1$. 
The corresponding p-value is obtained by comparing the value
of the observed test statistic ${\cal Q}$ with the simulated ${\cal Q}$ 
distribution for the background only hypothesis.
The post-trial p-value is $2.6\%$, which is equivalent to
2.2$\sigma$ (using the two-sided convention).

\begin{figure}[!htp]
\begin{center}
  \includegraphics[width=0.8\textwidth]{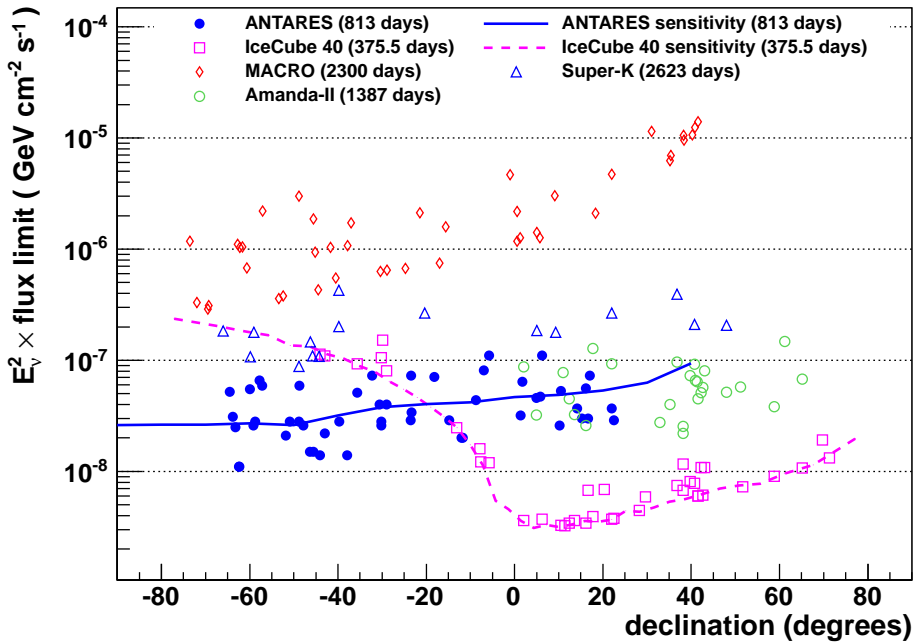}
  \caption{ Limits set on the $E^{-2}_{\nu}$ flux for the 51 sources in the candidate list (see Table~\ref{table:clresults}). Upper
    limits, previously reported by other neutrino experiments, on sources from both Northern and Southern sky are also included \citep{macro, SK, ic22}. The ANTARES sensitivity of this analysis is shown as a solid line and the IceCube 40 sensitivity as a dashed line \citep{icelim}.}
  \label{fig:limits}
\end{center}
\end{figure}

The results from the search up the 51 \textit{a priori} selected
candidate sources are presented in Table~\ref{table:clresults} and
shown in Figure~\ref{fig:limits}.  The most signal-like candidate 
source in the list is HESS J1023-575. The maximum likelihood fit yield  $\mu_{s} = 2.0$ and the test
statistic value is ${\cal Q} = 2.4$. The post-trial p-value of this
cluster is $41\%$, compatible with a background
fluctuation. Since no statistically significant cluster of events has
been found, upper limits (Feldman \& Cousins at $90\%$ CL) for an
$E^{-2}_{\nu}$ flux are obtained for the candidate list
sources. These limits and the corresponding $90\%$ CL sensitivity are plotted in Figure \ref{fig:limits} 
as a function of the source declination. Also indicated are the published limits from other experiments.

\begin{deluxetable}{cccccccccccc}
\tabletypesize{\scriptsize}
\tablecolumns{12} 
\tablewidth{0pc} 
\tablecaption{Results from the search for high-energy neutrinos from sources in the candidate list. The equatorial coordinates $(\alpha_{\rm s},\delta_{\rm s})$ in degrees, 
p-value ($\rm p$) probability and the $90\%$ C.L. upper limit on the $E^{-2}_{\nu}$ flux intensity $\phi^{90\% \rm CL}$ in units of $10^{-8} \rm GeV^{-1} cm^{-2} s^{-1}$ are
given (sorted in order of increasing p-value) for the 51 selected sources.}
\tablehead{ 
\colhead{Source name} & \colhead{$\alpha_s [^{\circ}]$}   & \colhead{$\delta_s [^{\circ}]$}  & \colhead{$\rm p$} & \colhead{$\phi^{90\%CL}$} 
&\colhead{   } &\colhead{   }  
&\colhead{Source name} & \colhead{$\alpha_s [^{\circ}]$}   & \colhead{$\delta_s [^{\circ}]$}  & \colhead{$\rm p$} & \colhead{$\phi^{90\%CL}$}}
\startdata 
HESS J1023-575  & 155.83  & -57.76 & 0.41 & 6.6  & & &  SS 433           & -72.04  & 4.98   & $-$ & 4.6 \\
3C 279          & -165.95 & -5.79  & 0.48 & 10.1 & & &  HESS J1614-518  & -116.42 & -51.82 & $-$ & 2.0 \\
GX 339-4        & -104.30 & -48.79 & 0.72 & 5.8  & & &  RX J1713.7-3946  & -101.75 & -39.75 & $-$ & 2.7  \\
Cir X-1         & -129.83 & -57.17 & 0.79 & 5.8  & & &  3C454.3         & -16.50  & 16.15  & $-$ & 5.5 \\
MGRO J1908+06   & -73.01  & 6.27   & 0.82 & 10.1 & & &  W28              & -89.57  & -23.34 & $-$ & 3.4  \\
ESO 139-G12     & -95.59  & -59.94 & 0.94 & 5.4  & & &  HESS J0632+057  &  98.24  & 5.81   & $-$ & 4.6  \\
HESS J1356-645  & -151.00 & -64.50 & 0.98 & 5.1  & & &  PKS 2155-304     & -30.28  & -30.22 & $-$ & 2.7  \\
PKS 0548-322    & 87.67   & -32.27 & 0.99 & 7.1  & & &  HESS J1741-302  & -94.75  & -30.20 & $-$ & 2.7 \\
HESS J1837-069  & -80.59  &  -6.95 & 0.99 & 8.0  & & &  Centaurus\ A     & -158.64 & -43.02 & $-$ & 2.1\\ 
PKS 0454-234     & 74.27   & -23.43 & $-$ & 7.0 & & &  RX J0852.0-4622 & 133.00  & -46.37 & $-$ & 1.5\\
IceCube hotspot  & 75.45   & -18.15 & $-$ & 7.0 & & & 1ES 1101-232     & 165.91  & -23.49 & $-$ & 2.8 \\
PKS 1454-354     & -135.64 & -35.67 & $-$ & 5.0 & & &  Vela X          & 128.75  & -45.60 & $-$ & 1.5\\ 
RGB J0152+017    & 28.17   & 1.79   & $-$ & 6.3 & & & W51C             & -69.25  & 14.19  & $-$ & 3.6\\
Geminga          & 98.31   & 17.01  & $-$ & 7.3 & & &  PKS 0426-380    & 67.17   & -37.93 & $-$ & 1.4\\
PSR B1259-63     & -164.30 & -63.83 & $-$ & 3.0 & & &  LS 5039          & -83.44  & -14.83 & $-$ & 2.7  \\  
PKS 2005-489     & -57.63  & -48.82 & $-$ & 2.8 & & &  W44             & -75.96  & 1.38   & $-$ & 3.1\\
HESS J1616-508   & -116.03 & -50.97 & $-$ & 2.7 & & &  RCW 86           & -139.32 & -62.48 & $-$ & 1.1  \\ 
HESS J1503-582   & -133.54 & -58.74 & $-$ & 2.8 & & &  Crab            & 83.63   & 22.01  & $-$ & 4.1\\
HESS J1632-478  & -111.96 & -47.82 & $-$ & 2.6  & & &  HESS J1507-622   & -133.28 & -62.34 & $-$ & 1.1  \\ 
H 2356-309       & -0.22   & -30.63 & $-$ & 3.9 & & &  1ES 0347-121    & 57.35   & -11.99 & $-$ & 1.9\\
MSH 15-52       & -131.47 & -59.16 & $-$ & 2.6  & & &  VER J0648+152    &  102.20 & 15.27  & $-$ & 2.8  \\ 
Galactic Center  & -93.58  & -29.01 & $-$ & 3.8  & & & PKS 0537-441    & 84.71   & -44.08 & $-$ & 1.3 \\
HESS J1303-631  & -164.23 & -63.20 & $-$ & 2.4  & & &  HESS J1912+101   & -71.79  & 10.15  & $-$ & 2.5  \\
HESS J1834-087   & -81.31  & -8.76  & $-$ & 4.3  & & & PKS 0235+164    & 39.66   & 16.61  & $-$ & 2.8 \\
PKS 1502+106    & -133.90 & 10.52  & $-$ & 5.2  & & &  IC443            & 94.21   & 22.51  & $-$ & 2.8  \\
                &         &        &           &     & & &  PKS 0727-11     & 112.58  & 11.70  & $-$ & 1.9 \\
\enddata 
\label{table:clresults}
\end{deluxetable}

\subsection{Limits for specific models}
Measurements of TeV gamma rays from the H.E.S.S. \citep{hess1, velax} telescope
may indicate a possible hadronic
scenario for the shell-type supernova remnant RX J1713.7-3946 and the
pulsar wind nebula Vela X. The first observation of RX J1713.7-3946 with 
the Fermi Large Area Telescope \citep{lat} shows that the gamma-ray emission 
seems to be compatible with a leptonic origin. However, composite models are also possible
as discussed in \citet{zika}.

\begin{figure}[!htp]
\begin{center}
  \includegraphics[width=0.50\textwidth]{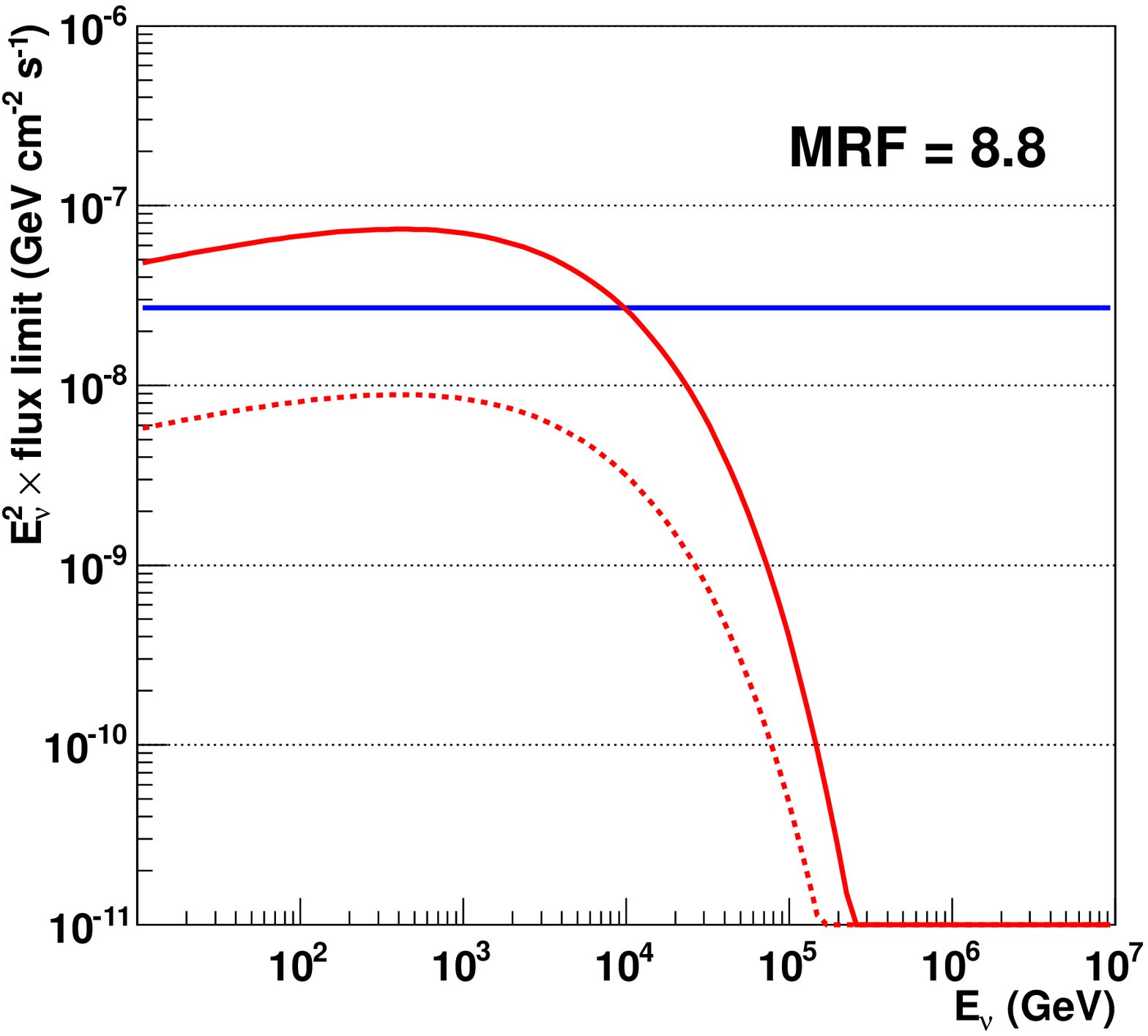}
  \includegraphics[width=0.48\textwidth]{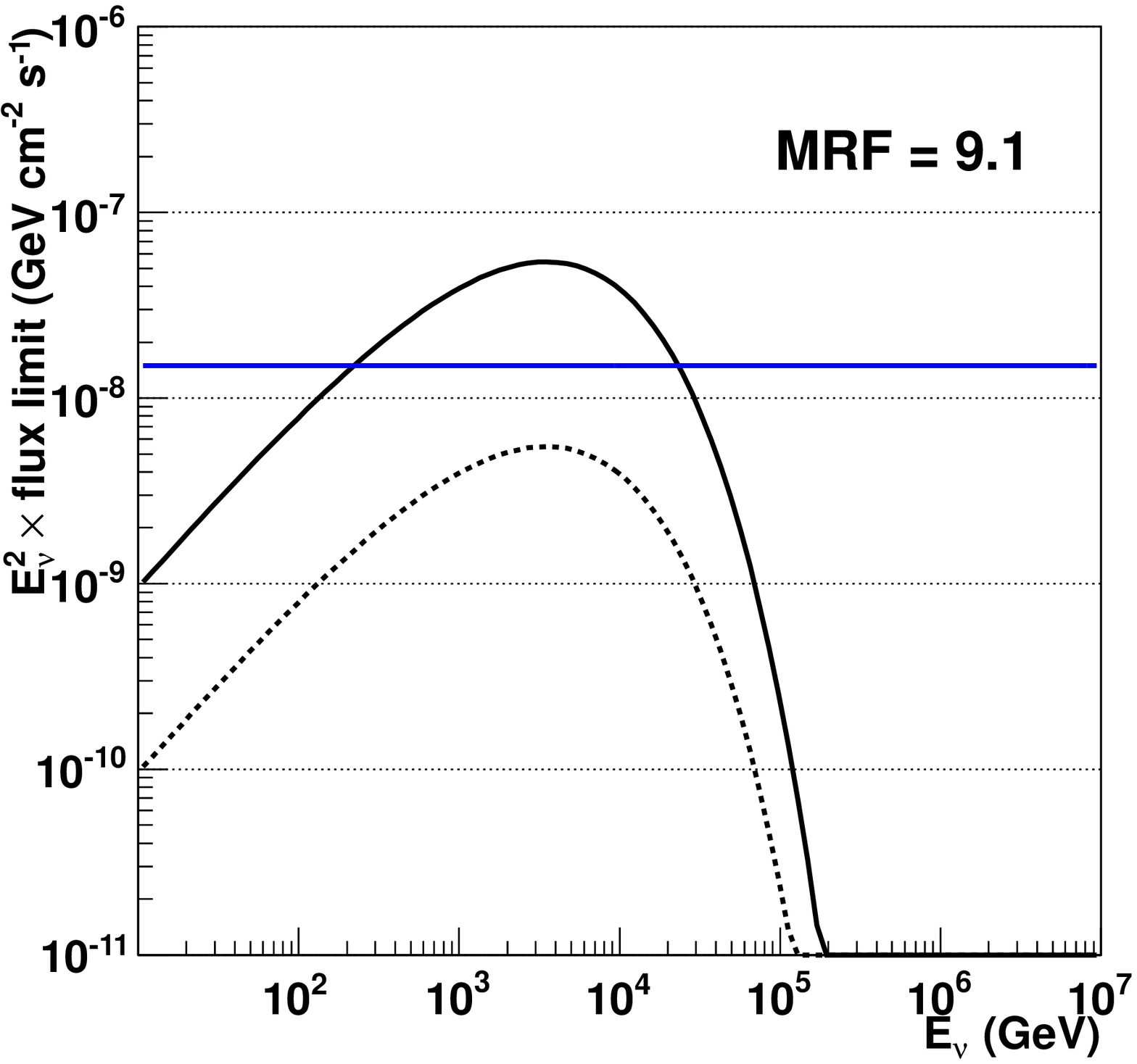}
  \caption{ Neutrino flux models (dashed lines) and 90\% CL upper limit (solid lines) for RX J1713.7-3946 (left) and Vela~X (right). Also shown are the $E^{-2}$ 
    point-source limit (solid horizontal blue lines) presented in Table~\ref{table:clresults}. 
    The models are taken from \citep{kappes}.}
  \label{fig:sources}
\end{center}
\end{figure}

In \citet{kappes} the neutrino flux 
and signal rates are estimated for these objects using 
the energy spectrum measured by
H.E.S.S and by approximating
the source extension with a Gaussian distribution. 
The spectrum for these models is shown in Figure~\ref{fig:sources}. 
Assuming these models, and taking into account the measured source extension, 
90\% CL upper limits on the flux normalisation were
computed for both sources. The Model Rejection Factor
(MRF) \citep{mrf}, i.e. the ratio between the 90\% CL upper limit and
the expected number of signal events, is also calculated.  Figure~\ref{fig:sources}
summarises these results. For RX J1713.7-3946 the upper limit is a factor
8.8 higher than the theoretical prediction.
For Vela X the upper limit is a factor 9.1 higher than the model. 
In both cases these are the most restrictive limits for the emission models considered.  

\section{Cross-check with the Expectation-Maximisation method}\label{sec:em}
The results discussed in the previous section have been cross-checked
using the Expectation-Maximisation (E-M) algorithm applied to the problem of the search for point
sources \citep{empaper, emps}. The E-M method is an iterative approach
to maximum likelihood estimations of finite mixture problems, which are
described by different probability density functions. In the case of a
search for point sources the mixture model can be expressed as the sum
of two components:

\begin{equation}
\label{eq_02}
\begin{split}
\rm{log} \mathcal{L}_{\rm s+b} &=\sum_{i} \rm{log}[(\frac{\mu_{\rm{s}}}{\mu_{\rm{t}}}) \times \mathcal{F}(\psi_{i}(\alpha_{\rm s},\delta_{\rm s})) \times \frac{\mathcal{N}^{\rm s}(N_{\rm hits}^{i})}{\mathcal{N}^{\rm bg}(N_{\rm hits}^{i})} \\
&+ (1-\frac{\mu_{\rm s}}{\mu_{\rm t}}) \times \mathcal{D}(\delta_{i})],
\end{split}
\end{equation}

\noindent
where the signal pdf ($\mathcal{F}$) is modelled as a two-dimensional
Gaussian, $\mathcal{D}$ is a polynomial parametrisation of the probability distribution
of the events in declination; as in Equation \ref{eq:lik}, the sum runs over all the
events in the data set, $\mu_{\rm t}$, and the number of hits is
used to better discriminate between background-like and signal-like events.

In comparison with the previous search method, the E-M algorithm uses a different likelihood description of the events 
and follows an analytical optimisation procedure that consists of two steps. In the {\it expectation} step the log-likelihood is
evaluated using the current set of parameters describing the source
properties. Then, during the {\it maximisation} step, a new set of
parameters is computed maximising the expected log-likelihood.  These
parameters are the number of signal events attributed to the source, the
source coordinates (in the full-sky search) and the standard
deviations of the Gaussian describing the signal pdf. In this sense
the E-M method has the freedom to adapt itself to the extension
of the source.  The test statistic used to determine the
significance of a potential point source is obtained as in Equation \ref{eq:test_q}.

\subsection{Results}
The most signal-like cluster found in the full-sky search is the same
as that found by the search method described in Section~\ref{sec:method}.
The number of signal events estimated by the algorithm is $\mu_{\rm
s}=5.3$. The observed value of the test statistic, ${\cal Q}=12.8$, or a
larger one occurs in $p=2.6\%$ of the background only experiments.
No significant excess of events was found in the location of any of
the candidate list sources. The lowest p-value is 0.87 (post-trial corrected)
and corresponds to the position of 3C-279. 
The results 
obtained with the two search methods described above are consistent.

\section{Conclusions}
\label{sec:conclusions}
The results of a search for cosmic neutrino point sources with the
ANTARES telescope using data taken in 2007-2010 have been presented. A
likelihood ratio method has been used to search for clusters of
neutrinos in the sky map. In addition to the position of the
reconstructed events, the information of the number of hits has been
used as an estimate of the neutrino energy. This improves the
discrimination between the cosmic signal and the background of
atmospheric neutrinos. 

Two searches have been performed: within a list
of candidate sources and in the whole sky. No statistically
significant excess has been found in either cases. In the full-sky search, the most
signal-like cluster is at ($\alpha, \delta$) = ($-46.5^{\circ}$,
$-65.0^{\circ}$). It consists of 9 events inside a 3 degrees cone, to
which the likelihood fit assigns 5.1 signal events. The corresponding
p-value is 0.026 with a significance of 2.2$\sigma$ (two-sided). The
most significant excess in the candidate list search corresponds to
HESS J1023-575, with a post-trial p-value of 0.41. 90\% CL Upper limits on the
neutrino flux normalisation are set at 1-10$\times$10$^{-8}$ GeV cm$^{-2}$ s$^{-1}$
in the range from 4 to 700 TeV (80\% of the signal), assuming
an energy spectrum of $E^{-2}_{\nu}$, and are the most restrictive ones for a large
part of the Southern sky. These limits are a factor $\sim$ 2.7 better
than those obtained in the previous ANTARES analysis based on the
2007-2008 data. Limits for specific models of RX J1713.7-3946 and Vela
X, which include information on the source morphology and spectrum, were also
given, resulting in a factor $\sim 9$ above the predicted fluxes.

\section*{Acknowledgements}

The authors acknowledge the financial support of the funding agencies:
Centre National de la Recherche Scientifique (CNRS), Commissariat
\`{a} l'\'{e}nergie atomique et aux \'{e}nergies alternatives  (CEA), 
Commission Europ\'{e}enne (FEDER fund and Marie Curie Program), 
R\'{e}gion Alsace (contrat CPER), R\'{e}gion
Provence-Alpes-C\^{o}te d'Azur, D\'{e}\-par\-tement du Var and Ville de
La Seyne-sur-Mer, France; Bundesministerium f\"{u}r Bildung und Forschung
(BMBF), Germany; Istituto Nazionale di Fisica Nucleare (INFN), Italy;
Stichting voor Fundamenteel Onderzoek der Materie (FOM), Nederlandse
organisatie voor Wetenschappelijk Onderzoek (NWO), the Netherlands;
Council of the President of the Russian Federation for young scientists
and leading scientific schools supporting grants, Russia; National
Authority for Scientific Research (ANCS - UEFISCDI), Romania; Ministerio
de Ciencia e Innovaci\'{o}n (MICINN), Prometeo of Generalitat Valenciana
and MultiDark, Spain; Agence de l'Oriental and CNRST, Morocco. We also
acknowledge the technical support of Ifremer, AIM and Foselev Marine
for the sea operation and the CC-IN2P3 for the computing facilities.

\end{document}